\documentclass[a4paper,11pt]{article}
\pdfoutput=1 % if your are submitting a pdflatex (i.e. if you have
\usepackage{jcappub} % for details on the use of the package, please
\usepackage[T1]{fontenc} % if needed

\title{\boldmath Constraining cosmic polarization rotation and implications for primordial B-modes}

\author[a]{Joel Williams,}
\author[a]{Aditya Rotti,}
\author[a]{Richard Battye}

\affiliation[a]{Jodrell Bank Centre for Astrophysics, Department of Physics and Astronomy, School of Natural Sciences, University of Manchester, Manchester, M13 9PL, U.K.}

\emailAdd{christopher.williams-4@manchester.ac.uk, aditya.rotti@manchester.ac.uk, richard.battye@manchester.ac.uk}

\abstract{Cosmological Birefringence (CB) is a phenomenon, caused by parity violating modifications to electrodynamics, whereby the linear polarisation angle of light changes as photons traverse a vacuum. It is possible to use a number of different analysis techniques to constrain this effect using Cosmic Microwave Background (CMB) polarisation observations. We investigate two different methods of constraining direction dependent birefringence for present and future CMB experiments including BICEP/Keck, Simons Observatory (SO), and \emph{LiteBIRD}. Specifically we compare the constraints placed on anisotropic CB from a quadratic estimator technique to those derived from estimates of the $B$-mode power-spectrum for the three different experiments. The constraints derived from estimates of the $B$-mode power spectrum are found to be comparable to those derived from quadratic estimator for BICEP/Keck and SO, but not \emph{LiteBIRD} due to its larger sky coverage. These forecasted upper bounds for CB are converted to constraints on primordial magnetic fields and the coupling between photons and pseudo Nambu-Goldstone bosons. Finally we show that even with the best constraints on CB, for the respective experiments, the potentially induced $B$-mode power can act as a significant contaminant in the prospective measurement of primordial $B$-modes.}

\begin{document}
\maketitle
\flushbottom

\section{\label{sec:Introduction}Introduction}

\noindent The temperature and polarization anisotropies in the Cosmic Microwave Background (CMB)\cite{Penzias1965} have been measured to high precision by the Wilkinson Microwave Anisotropy Probe (WMAP) and \textit{Planck} satellite, placing very tight constraints on the cosmological parameters \citep{Hinshaw_2013, Planck2015, Planck2018params}. These observations have confirmed the standard $\Lambda$CDM cosmology, and are compatible with the predictions of inflation. The current observations (the temperature anisotropies, E-modes of polarization and B-modes sourced by weak lensing) primarily probe the scalar density perturbations. Most inflation models also predict the generation of tensor perturbations (gravitational waves). However, the predicted range of amplitudes, characterized typically by the ratio of power in tensor perturbations to power in scalar perturbations `$r$', has no lower bound. These tensor perturbations source very specific B-mode of polarization patterns, with a fairly well known spectral shape,  which has the most power on degree scales around $\ell \sim 80$. The ongoing (BICEP/Keck, Spider, PolarBear, ACT, SPT \cite{PhysRevLett.114.101301,Gualtieri2018,Ade2014,Sievers_2013,Schaffer_2011}) and upcoming CMB (CMB) experiments (Simons Observatory (SO), \emph{LiteBIRD}, CMB-S4 \cite{Ade_2019, Hazumi2019, Abazajian2019})  will measure  the polarized CMB sky with an unprecendented precision and with exquisite control over systematics over the next decade; a robust detection of $r$ being one of their primary science goals. These experiments are projected to improve the upper limits on r from the current limit of $r<0.061$ \cite{collaboration2018planck} to $r\lesssim10^{-3}$.
\par
This program of research assumes that $B$-modes are only produced by inflationary gravitational waves. However, there may be other non-standard mechanisms which also generate B-modes. If these alternate sources of B-modes exist, then these could potentially act as contaminants to measurements of B-modes specifically induced by inflationary tensor perturbations and it is important to develop analysis techniques which allow us to distinguish between these different sources.
\par 
A well known example is that of B-mode generation due to weak gravitational lensing of the CMB\cite{LEWIS20061}. Weak lensing results in the subtle remapping of the microwave polarization vectors, which in effect leaks some of the E-mode power into B-modes. To detect $r\sim 10^{-3}$ it is imperative to ``de-lens'' the polarized sky in order to separate the primordial B-modes from those generated due to weak lensing. Consequently, a significant effort has been invested to carry out robust  ``de-lensing'' \cite{PhysRevD.62.043007, PhysRevD.90.023539, PhysRevD.92.043005,Carron_2017}.

In this work, however, we focus on an alternative mechanism: B-modes generated due to cosmic polarization rotation (CPR) which could be sourced by a so called Cosmological Birefringence (CB) field \cite{Carroll1990, Carroll1998}. While we use this as our primary motivation to carry out this study, the constraints and methods discussed in this work are also applicable to other mechanism that results in a rotation of the CMB polarization. 

\subsection{Modified Electrodynamics}
\par
\noindent Parity violations are common place in the weak sector of the standard model, with many observational examples. However, the electromagnetic sector of the standard model as it is currently understood is not expected to contain parity violating interactions \cite{Lue1999}. Despite this, the search to understand the dark sector and inflation has introduced a plethora of potential pseudo-scalar fields, such as the axion \cite{Marsh2016}. Such a pseudo-scalar is a Pseudo-Nambu-Goldstone Boson (PNGB) which can can couple to the gauge field through a Chern-Simons term \cite{Carroll1990, Carroll1998, Leon2017,Frieman1995}
\begin{equation}\label{eqn:chern simons lagrangian}
    \mathcal{L}_{CS} = -\frac{\beta \phi}{2M}F_{\mu\nu}\tilde{F}^{\mu\nu}\;,
\end{equation}
where $\beta$ is a dimensionless coupling constant, $M$ is the vacuum expectation value of the broken global symmetry, $F_{\mu\nu}$ is the electromagnetic field strength tensor and $\tilde{F}_{\mu\nu}$ is its dual.
\par 
In this case the parity violating physics introduced by the additional Chern-Simons term induces a difference between the effective refractive indexes for the right-handed and left-handed circular polarisation states of light. Linearly polarised light can be written as the superposition of the two circular polarisation states with a polarisation angle proportional to the difference in phase between the two states. The resulting change in phase during the propagation of light, due to difference in the effective refractive index for the different circular polarization states will cause a change in the linear polarisation angle \cite{Carroll1990}. This effect is known as cosmological birefringence and the resulting change in polarisation angle in direction $\hat{\mathbf{n}}$ is \cite{Caldwell2011}
\begin{equation}
    \bar{\alpha}(\hat{\mathbf{n}}) = \frac{\beta}{M}\int d\eta\left(\frac{\partial}{\partial\eta} + \hat{\mathbf{n}}\boldsymbol{\cdot}\boldsymbol{\nabla}\right)\phi(\eta,\hat{\mathbf{n}})\;,
\end{equation}
where $\eta$ conformal time and $\bar{\alpha}(\hat{\mathbf{n}}) = \alpha_{0} + \alpha(\hat{\mathbf{n}})$. Here the integration is along the space-time path of the photon. Note, that this angle is direction dependent only if the gradient of the scalar field is also spatially varying. Otherwise the rotation angle is direction independent, $\bar{\alpha}(\hat{\mathbf{n}})=\alpha_{0}$. The power spectrum for this spatially dependent field is expected to have a simple form,
\begin{equation}\label{eqn:Birefringence Power spectrum}
    C^{\alpha\alpha}_{L} = A_{CB}\frac{2\pi}{\left[L(L+1)\right]^{\beta}}\;,
\end{equation}
where the constant $A_{CB}$ is the amplitude of the power spectrum, a parameter which CMB experiments might seek to constrain.  It is assumed that as long as the pseudo-scalar field that sources CB does not obtain a mass during the inflationary epoch then the power spectrum will be scale invariant, corresponding to $\beta=1$ \cite{Caldwell2011}. While some models predict oscillatory damping of the CB spectrum at high multipoles \cite{Zhao2014,Capparelli2020}, this scale invariant assumption is expected to remain valid in the spectral ranges probed by the experiments considered in this analysis. The true spectral shape of the CB spectrum remains unknown, so we have allowed for perturbation around scale invariance by including the additional free parameter $\beta$.
\par

\par
\subsection{Birefringence and the CMB}
\noindent The CMB polarisation anisotropies were generated in the Early Universe during the epoch of recombination via Thomson scattering. Approximately $10\%$ of these CMB photons are linearly polarised. The high redshift origin of these polarised photons makes the CMB an ideal candidate for the detection of CB. 
\par 
CB affects CMB polarization maps in a manner analogous to weak lensing modifications to the maps. Therefore CB not only modifies the angular power spectra (i.e. the diagonal of the harmonic space covariance matrix : $\langle a^X_{\ell m} a^Y_{\ell' m '} \rangle$) but also encodes information in the off-diagonal elements of the covariance matrix. The monopole component of the CB rotation angle causes mixing between the different angular power spectra at first order in $\alpha_0$ - the uniform CB rotation angle. The anisotropies in the CB rotation angle also modify the different angular power spectra but leading order corrections appear only at second order~($\alpha ^2$). The anisotropies in the CB rotation angle are also encoded in the off-diagonals of the covariance matrix, and these are at first order~($\alpha$) and these therefore can be reconstructed using the well known quadratic estimator (QE) technique \cite{PhysRevD.92.123509,Yadav2012,Gluscevic2009,Kamionkowski2009}.
\par
It is therefore possible to put constraints on CB affects in CMB polarization maps using two complementary methods. One can look for excess power in direct measurement of the polarization angular power spectra. The details of the modification to the polarization angular power due to CB are summarized in Section.~\ref{sec:math_ps}. Alternately, one can use the QE technique, where $\alpha (\hat{n})$ is reconstructed, and the power spectrum of the reconstructed map is compared to the null hypothesis of it being consistent with noise. Relevant details relating to the QE are discussed in Section.~\ref{sec:math_qe}.
It is important to note that constraining uniform birefringence is limited by the level of absolute polarisation angle calibration possible \cite{Kaufman:2014rpa}, however, such limitations do not apply when placing constraints on anisotropic birefringence. In this work we assume that any monopole birefringence effects have already been removed using self calibration \cite{Keating_2012} and focus on deriving constraints on anisotropic birefringence from upcoming CMB experiments using the two complementary methods discussed above. In Section.~\ref{sec:likelihood} we discuss the likelihoods we use, and the experimental configurations for BICEP/Keck, SO and \emph{LiteBIRD} to derive constraints for each of these two methods.
\par
Many CMB experiments have already placed constraints on both uniform and anisotropic CB \cite{PhysRevD.92.123509,LIU201722,Contreras_2017,Ade2017,Gluscevic2012,Kaufman:2014rpa}. The current best constraint on the uniform CB rotation angle, $\alpha_{0}$, comes from \emph{Planck} whose 68\% confidence limit is $\alpha_{0} < 0.5^{\circ}$ \cite{refId0}. The best constraint on  anisotropic birefringence is a 95\% confidence limit constraint on the amplitude of the CB power spectrum of $A_{CB} \leq 0.10\times10^{-4}\,{\rm rad}^{2}$, set by analysis on recent ACT, assuming a scale invariant power spectrum for CB~(i.e.  $\beta=1$) \cite{Namikawa2020}. In Section.~\ref{sec:Results} we present a forecast of the CB constraints from BICEP/Keck, SO and \emph{LiteBIRD}. A qualitatively similar study was carried out in \cite{Pogosian_2019} and we find consistent results under similar settings. Further to the presentation of this forecast we compare the relative constraints one may be able to obtain using both the QE technique, and by looking for excess power in the polarization angular power spectra.
\par 
Due to the relationship between the CB and the coupling strength between the photon and the PNGB it is possible to forecast constraints for this coupling using constraints on CB. As the observable for CB, rotation of the CMB linear polarization angle, is the same as the observable for primordial magnetic fields (PMFs) it is possible to also use CB constraints to place constraints on the field strength of PMFs \cite{PhysRevD.86.123009,PhysRevD.88.063527,10.1093/mnras/stt2378}. Forecasts for constraints on both physical phenomenon for BICEP/Keck, SO, and \emph{LiteBIRD} are presented in Section.~\ref{sec:Results}. 
\par 

A measurement of the primordial B-modes, originating from tensor perturbations to the metric as predicted by most models of inflation, is one of the primary science goals of many observational programs. Since CB can induce excess B-mode power which can be potentially confused with these primordial B-modes it is important to understand the constraints on CB attainable via different experiments. This will be an important aspect of interpreting the B-mode measurements of the future. With this motivation, in Section~\ref{sec:Results} we compare the CB induced $B$-mode allowed by the forecasted upper bounds on CB for BICEP/Keck, SO small aperture telescope (SAT) and large aperture telescope (LAT), and \emph{LiteBIRD}. An analogous theoretical study was performed in \cite{Zhao2014}, where the was focus was on estimating induced B-modes sourced by CB spectra corresponding to different coupling strengths between a PNGB and CMB photons. We emphasize that here we estimate the B-mode power that cannot be ruled out, even after taking into account the best upper limits on CB that will be placed by the corresponding experiments\footnote{Here we implicitly assume that no significant detection of CB is made by the corresponding experiment.}.

\section{\label{sec:maths}Constraining CB using CMB polarisation}
\noindent The affects of anisotropic CB on the CMB polarisation are expected to be small and hence can be treated perturbatively. The CMB polarization is written in terms of the maps of the Stokes parameters $Q(\hat{\mathbf{n}})$ and $U(\hat{\mathbf{n}})$ on the sky. Defining the complex Stokes parameters,
\begin{equation}
    {}_{\pm}P(\hat{\mathbf{n}}) \equiv (Q\pm iU)(\hat{\mathbf{n}})\;,
\end{equation}
the rotation of the polarisation due to birefringence is given by: ${}_{\pm}P(\hat{\mathbf{n}}) = {}_{\pm}\tilde{P}(\hat{\mathbf{n}})e^{\mp i2\alpha(\hat{\mathbf{n}})}$. Note that tilde'd variables are used to denote the primordial (un-rotated) CMB fields. We reiterate that we focus only on anisotropic CB and assume that the monopole CB (or angle miscalibration) has already been corrected using self calibration. Treating this perturbatively and retaining terms to second order in $\alpha$ yield the following correction to the polarization vector,
\begin{equation}
\label{eqn:alpha perturbation}
    {}_{\pm}P(\hat{\mathbf{n}}) = {}_{\pm}\tilde{P}(\hat{\mathbf{n}}) \left[ 1  \mp i2\alpha(\hat{\mathbf{n}}) \pm 2\alpha^2(\hat{\mathbf{n}}) + \mathcal{O}(\alpha^3)\right]\,.
\end{equation}

The E and B fields are an equivalent, but scalar (spin 0) representation of CMB polarization. In the following section we summarize how these CB corrections propagate to the harmonic space covariance of the scalar E and B fields.
%%%%%%%%%%%%%%%%%%%%%%%%%%%%%%%%%%%%%%%%%%%%%%%%%%%%%%%%%%%
\subsection{The effect of CB on the B-mode power-spectrum}
\label{sec:math_ps}

\noindent As previously mentioned, the dominant corrections to the CMB polarization angular power spectra, due to anisotropic CB rotation angle, appear at second order in $\alpha$. These corrections result in mixing of power between different multipoles and also in mixing of power between the E and B mode of polarization. 

The correction to the angular power spectrum of B-mode of polarization is given by \cite{Li2015},
\begin{equation}
    \delta C_{\ell}^{BB}=\frac{1}{\pi}\sum_{L}C^{\alpha\alpha}_{L}(2L+1)\sum_{l_{2}}(2l_{2}+1)\tilde{C}^{EE}_{l_{2}}(H^{L}_{ll_{2}})^{2} + \mathcal{T}_B(B \rightarrow B)\,,\label{eqn:deltaclbb}
\end{equation}
where only modes that satisfy the triangularity condition $l+L+l_{2}=\text{Even}$ contribute to the sum and $H^L_{ll'} = \left(\begin{array}{ccc}
         l & L & l'  \\
         2 & 0 & -2  
    \end{array}\right)$ is a Wigner symbol. 
Similarly the correction to the angular power spectrum of the E-mode of polarization has the following form \cite{Li2015},
\begin{equation}
    \delta C_{\ell}^{EE} =\frac{1}{\pi}\sum_{L}C^{\alpha\alpha}_{L}(2L+1)\sum_{l_{2}}(2l_{2}+1)C^{EE}_{l_{2}}(H^{L}_{ll_{2}})^{2} +\mathcal{T}_E(B \rightarrow E) \,, \label{eqn:deltaclee}
\end{equation}
where only modes that satisfy the triangularity condition $l+L+l_{2}=\text{Odd}$ contribute to the sum. For brevity the explicit form for additional corrections $\mathcal{T}_B$ and $\mathcal{T}_E$ are not given here, since these are sub-dominant. Unlike in the case of monopole birefringence, the anisotropic birefringence does not generate $C_l^{EB}$ when including corrections up to second order in $\alpha$.

Since the power in the E-modes is significantly larger than in the B-modes, the dominant corrections result from mixing of E-mode power across multipoles and across polarization states. This is also the reason why the additional corrections ($\mathcal{T}_B$ \& $\mathcal{T}_E$) sourced by B-mode power are sub-dominant. On evaluating the correction to the CMB polarization power spectrum (Eq.~\ref{eqn:deltaclee} \& Eq.~\ref{eqn:deltaclbb}), with $C_{L}^{\alpha \alpha}$ consistent with the current constraints on anisotropic CB, it is seen that the corrections to the $E$-mode power spectrum are more than an order of magnitude below cosmic variance. This indicates that the E-mode power spectrum cannot place interesting constraints on CB. On the contrary corrections to the B-mode power spectrum are comparable to primordial B-mode power expected from the range of tensor to scalar ratio being targeted by current and upcoming CMB experiments. This indicates that the measured B-mode power spectrum can be used to place interesting constraints on CB.
\par

\subsection{The Quadratic Estimator}
\label{sec:math_qe}
\noindent As noted previously, CB induces coupling between off-diagonal elements of the covariance matrix\footnote{In the absence of CB the off-diagonal correlations are zero, except when considering analogous couplings induced by weak lensing of the CMB.}, at leading order in $\alpha$, that can be measured from data and combined in an optimal manner, using the (QE) technique, to yield a map of $\alpha$. This is entirely analogous to the more well known case of QE reconstruction of the weak lensing potential \cite{PhysRevD.67.083002}. While the details of constructing these QE for CB can be found in \cite{Gluscevic2009,Kamionkowski2009}, here we summarize the QE details relevant to this work.

In practice, one carries out the CB reconstruction by extracting information from the each of the following covariance matrices: EB, EE, BB, TE \& TB and finally combining the reconstruction from each estimator in an optimal way, duly accounting for the correlations between the different estimates. However it is seen that the final result is dominated by the EB QE and the explicit form of this estimator is given by,

\begin{equation}\label{eqn:full quadratic estimator}
    \widehat{\alpha}_{LM}= -2N_{L}\sum_{ll'}\frac{\tilde{C}^{EE}_{l'}}{C^{BB}_{l}C^{EE}_{l'}}\sum_{mm'}B_{lm}E^{*}_{l'm'}\xi^{LM}_{lml'm'}\;,
\end{equation}
where $\tilde{C}_{l}$ represents the true primordial CMB power spectrum that is calculated using a Boltzmann code, such as CAMB, with an assumed fiducial cosmology, $C_{l}$ denotes the observed power spectrum: $C_{l} = \tilde{C}_{l}+C^{\rm noise}_{l}$, $E_{lm}$ and $B_{lm}$ are beam deconvolved harmonic space fields, $\xi^{LM}_{lml'm'}$ represents the geometric kernel that is given by the following expression,
\begin{equation}\label{eqn:xi definition}
     \xi^{LM}_{lml'm'} = (-1)^{m}\sqrt{\frac{(2l+1)(2L+1)(2l'+1)}{4\pi}}\left(\begin{array}{ccc}
          l & L & l'  \\
          -m & M & m' 
     \end{array}{}\right) H^{L}_{ll'}\;,
\end{equation}

and $N_L$ is the reconstruction noise power spectrum. $N_L$ is the expected power spectrum of the reconstructed map in the absence of any CB and is given by the following expression,
\begin{equation}\label{eqn:full reconstruction noise}
    N_{L} = \left[4\sum_{ll'} \sqrt{\frac{(2l+1)(2l'+1)}{4 \pi}}\frac{(\tilde{C}^{EE}_{l'}H^{L}_{ll'})^{2}}{C^{BB}_{l}C^{EE}_{l'}} \right]^{-1}\;.
\end{equation}
It is important to note that, owing to parity conditions for the EB QE estimator, the sums in the estimator and the reconstruction noise only receive non-zero contribution when the condition $l+l'+L=\text{Even}$ is satisfied. To make forecasts using the QE technique, one only needs to evaluate $N_L$ using Eq.~\ref{eqn:full reconstruction noise}, which requires the fiducial CMB power spectra, the noise and instrument beam for each experiment as an input. 

\section{Forecast methodology for $A_{CB}$\label{sec:likelihood}}

We employ a likelihood based approach to constrain the amplitude of the CB power spectrum, $A_{CB}$, using both the QE approach, and using direct observations of the $B$-mode power spectrum (BB). We forecast the upper bound on $A_{CB}$ using both these QE and BB techniques for SO, \emph{LiteBIRD}, and a simulated version of BICEP/Keck. As a robustness check we also derive $A_{CB}$ limits using QE and BB techniques using actual BICEP/Keck band-power data and compare them against those quoted by BICEP/Keck \cite{Ade2017}. 
\par
For these analyses we work with a null hypothesis for primordial $B$-modes (i.e. $r=0$). The maximum multipole is fixed to the same value for all experimental configurations. By so doing, the different multipole contributions to the likelihood are naturally determined by the noise in the respective measurements. 
The likelihood is given by $\mathcal{L}= \mathcal{N}e^{-\frac{1}{2}\chi^{2}}$ and the $\chi^{2}$ has the following general form,
\begin{eqnarray}\label{eqn:comparison likelihood}
    \chi^{2}(A_{CB}) =&&\, \Delta d_l M_{ll'}^{-1} \Delta d_l'\;,
\end{eqnarray}
where $\Delta d_l = \left[C^{\rm obs}_{l} - C^{\rm model}_{l}(A_{CB})\right]$ and $M_{ll'}$ is the covariance matrix evaluated at a fixed fiducial cosmology. Note that even for the model power spectra $A_{CB}$ is the only parameter that is allowed to vary while other cosmological parameters are held fixed. The CB power spectrum is assumed to be a power law (see Eq.\ref{eqn:Birefringence Power spectrum}) and we study the constraints by allowing the slope to vary by 15\% around a scale invariant power spectrum.

In the following sub-sections we provide the specifics of the model spectra, observed spectra and covariance matrices used to define the BB and QE likelihoods.
\par 

\subsection{The BB likelihood for $A_{CB}$}

As discussed in Section~\ref{sec:math_ps} a non-vanishing CB effect will result in excess B-mode power, analogous to how weak lensing generates B-mode power by leaking some of the E-mode power. It is possible to put constraints on $A_{CB}$ by searching for this excess power in the measured B-mode power spectrum. To do this we use the following definition of the model spectrum to evaluate the likelihood function \cite{PhysRevD.92.123509},

\begin{equation}\label{eqn:BB comparison likelihood}
 C^{\rm model}_{\ell}(A_{CB}) = A_{CB}C^{\rm BB, CB}_{\ell} + A_{\rm lens}C^{\rm BB, lens}_{\ell}\,,
\end{equation}
where $C^{\rm BB, CB}_{\ell}$ is the CB induced BB power spectrum which is evaluated by injecting the CB power spectrum, given in Eq.~\ref{eqn:Birefringence Power spectrum}, into Eq.~\ref{eqn:deltaclbb} for $A_{CB}=1$ and a range of values of $\beta$.
The covariance matrix, $M_{\ell\ell'}$, then has the following form,

\begin{equation}
    M_{\ell\ell'} = \frac{2}{(2\ell+1)f_{\rm sky}}\left[A_{\rm lens}C^{BB,\rm lens}_{\ell} + C^{BB,\rm noise}_{\ell}W_{\ell}^{-2}\right]^{2}\;,
\end{equation}

where, $C_{\ell}^{BB,\,\rm lens}$ is the $B$-mode power spectrum induced by lensing, $A_{\rm lens}$ generally characterizes the lensing power post de-lensing, $C^{BB,\rm noise}_{\ell}$ is the instrument noise power spectrum and $W_{\ell}$ is the beam window function. When using this method we also derive constraints for different amounts of de-lensing, which is done by simply using different values, $A_{\rm lens} \in \{1,0.75,0\}$\footnote{$A_{lens}=1$ means no de-lensing, while $A_{lens}=0$ means perfect de-lensing.}, as this considerably changes the effective noise in the measurements of the BB power spectrum.
 
\subsection{The QE likelihood for $A_{CB}$\label{sec:qe likelihood}}
In addition to inducing excess B-mode power, CB would also induce specific signatures in the off-diagonal elements of the covariance matrix and these can be used to reconstruct the CB rotation field, $\alpha$, as discussed in Section \ref{sec:math_qe}. For the QE likelihood $C_l^{obs}$ would be the power spectrum of the reconstructed $\alpha$ map which we denote by $C^{\alpha\alpha,\,\rm rec}_{L}$. The model spectrum is then given by \cite{Ade2017},

\begin{equation}\label{eqn:QE comparison likelihood}
C^{\rm model}_{L}(A_{CB}) = A_{CB}C^{\alpha\alpha,\,\rm ref}_{L} +  N_{L} \;,
\end{equation}
where $C^{\alpha\alpha,\,\rm ref}_{L}$ denotes a reference CB power spectrum and is assumed to have the same form as in. Eq. \ref{eqn:Birefringence Power spectrum}. $N_L$ denotes the reconstruction noise power spectrum which can be evaluated using Eq.~\ref{eqn:full reconstruction noise}. Note that $N_L$ can be evaluated given only the theoretical CMB spectra and the instrument noise power spectrum. The covariance matrix $M_{ll'}$ for the QE likelihood has the following form,

\begin{equation}
    M_{LL'} = \frac{2}{(2L+1)f_{\rm sky}} N_{L}^{2}\;.
\end{equation}

Unlike in the BB analysis where we derive constraint for different values of $A_{\rm lens}$, here we keep $A_{\rm lens}=1$ fixed.
We estimate the reconstruction noise by optimally combining the reconstruction noise for all the QE (EE, BB, EB, TE, TB) estimators. For all experimental configurations we find that this net reconstruction noise offers little improvement on $A_{CB}$ constraints compared to those derived from using the reconstruction noise corresponding to the $EB$ estimator alone.
 
\subsection{\label{Experimental}Experimental configurations}
The multipole ranges, sky coverage, noise and beam for the three experiments considered in this work are summarised in Table~\ref{tab:Instrument inputs}. Instead of artificially varying the $l_{\rm max}$ cutoff for each experiment, the maximum multipole of $l_{\rm max}=3000$ was chosen so that the cutoff is instead where the signal becomes saturated by noise for each experiment.

\begin{table}[h!]
    \centering
    \begin{tabular}{cccccc}
    \hline\hline
        Instrument & $\ell_{\rm min}$ & $\ell_{\rm max}$ & $f_{\rm sky}$ & noise rms &$\Theta_{\rm FWHM}$  \\
        &&&&[$\rm\mu K$arcmin]& [arcmin]\\
        \hline\hline
         BICEP/Keck & 30 & 3000 & 0.01 & 3.0 & 30\\
         
         SO SAT 93GHz & 30 & 3000 & 0.1 & 3.8 & 30\\
         
         SO LAT 93GHz & 30 & 3000 & 0.4 & 16.3 & 2.2\\
         
         \emph{LiteBIRD} & 2 & 3000 & 0.7 & 2.5 & 30\\
         \hline\hline
    \end{tabular}
    \caption{Instrument specifications used in construction of the respective likelihood functions. For BICEP/Keck, noise  curves were fit to publicly available noise data. For SO the noise curves were produced using a publicly available SO noise forecast code. For both SO and BICEP/Keck the noise rms presented is the average value for these noise curves in the range $30 \leq \ell \leq 3000$. The noise curves for LiteBIRD were calculated directly from the beam and noise rms values.}
    \label{tab:Instrument inputs}
\end{table}

The noise power spectrum used and the instrument beams for each analysis carried out in this work were chosen with the following prescriptions/reasons:
\begin{itemize}
    \item BICEP/Keck actual band-power data: The publicly available 150GHz binned noise data was chosen \cite{Ade2017} and the model spectra were identically binned using the prescription presented in \cite{Hivon_2002}. 
    \item BICEP/Keck simulated forecast: The Gaussian noise model was fitted to the publicly available 150GHz channel binned noise\footnote{The BICEP/Keck binned noise and band power data is available at \url{http://bicepkeck.org/bk15_2018_release.html}.}\cite{Ade2017}. We ensured that this fitted noise model closely resembles the BICEP noise. This procedure allows us to use an extended multipole range for making a forecast. The limit of $\ell_{\rm min} = 30$ for the simulated forecast was chosen to emulate the baseline QE analysis performed in \cite{Ade2017}.
    \item SO forecast: A publicly available noise curve code\footnote{The noise curves for the SO LAT and SAT telescopes can be found at \url{https://simonsobservatory.org/assets/supplements/20180822_SO_Noise_Public.tgz}.} was used to generate  noise curves for the 93GHz  channel. This channel was chosen as it gives the strongest constraint on $A_{CB}$ \cite{Ade_2019}. Both the SAT and LAT noise curves were generated using the ``baseline" mode of the noise code \footnote{The ``one over $f$" mode in the noise code was set to ``optimistic"}. In addition to instrument noise, these noise curves include additional contributions from atmospheric noise. While for forecasts, we use these simulated noise curves, the rms noise quoted in Table~\ref{tab:Instrument inputs} are estimated from fitting the amplitude of the Gaussian noise model to the simulated noise curved. The value of $\ell_{\rm min} = 30$ was chosen in order to be consistent with the $\ell$ range used for the SO forecasts presented in \cite{Ade_2019}. 
    \item \emph{LiteBIRD} forecast: A noise curve was constructed using the average beam width and the rms noise values quoted in \cite{Hazumi2019}.
\end{itemize}
\section{\label{sec:Results}Results}
\subsection{Forecasted constraints on $A_{CB}$}
\begin{figure*}[t!]
    \centering
    \includegraphics[width=1\textwidth]{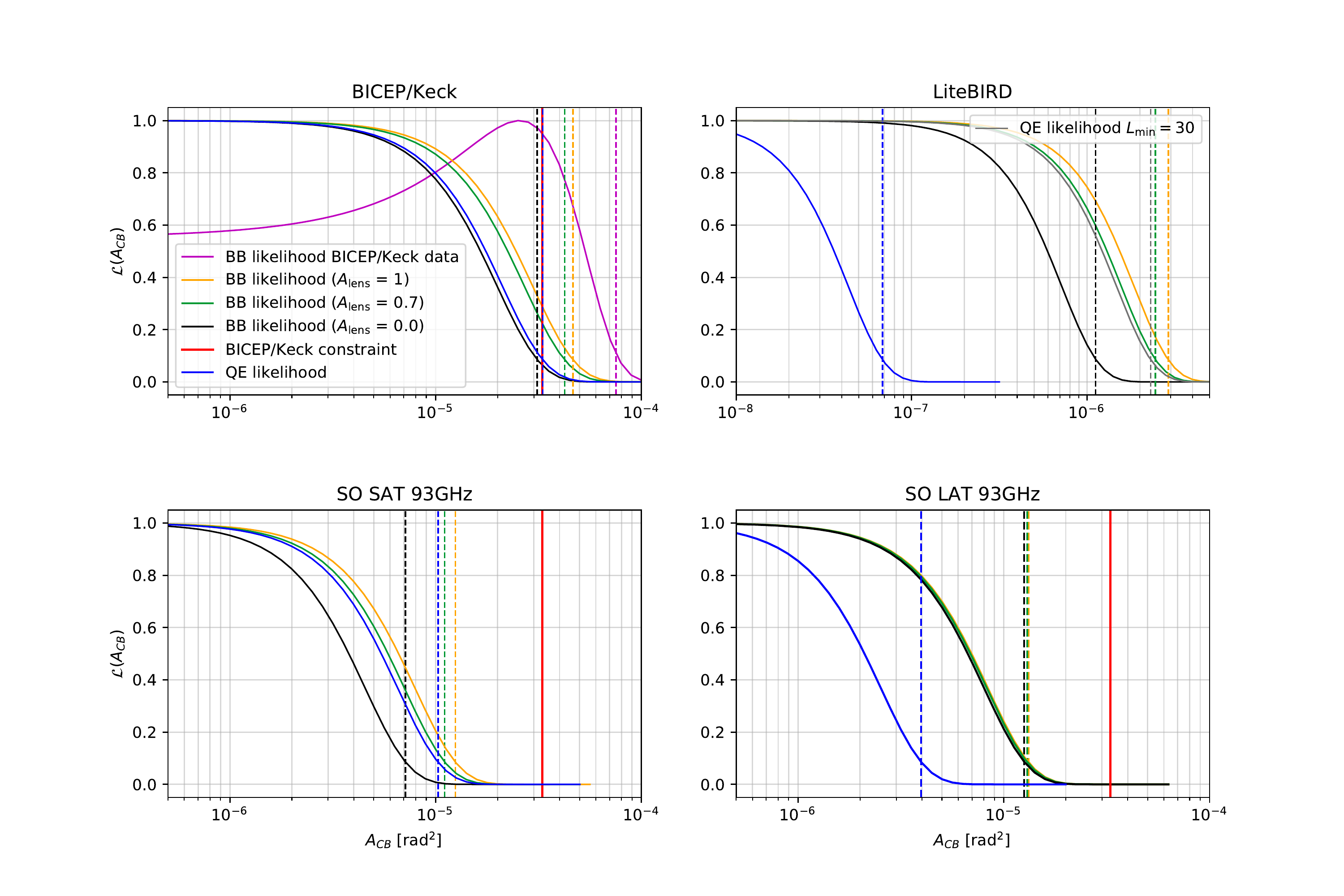}
    \caption{Forecasted likelihoods and 95\% confidence limits for $A_{CB}$ for the LAT and SAT SO 93GHz channels, BICEP/Keck, and \emph{LiteBIRD}. The current experimental upper bound on $A_{CB}$ of  $A_{CB} < 0.33\times10^{-4}\;\text{rad}^{2}$from \cite{Ade2017} is indicated as the red vertical line. The forecasted constraint from the QE is shown in blue, whereas the forecast constraint from BB when no de-lensing is present is indicated in orange. The expected constraints from BB when $A_{\rm lens} = 0.7$ is indicated in green, and for perfect de-lensing ($A_{\rm lens} = 0$) is indicated in black. The constraint from the actual BICEP/Keck $B$-mode bandpower data is indicated in magenta. For the reference spectra used to calculate both the QE and BB likelihood curves shown $\beta=1$. For the ground based experiments, BICEP/Keck and SO, the multipole range for the forecast is $30\leq \ell \leq3000$ and $30\leq L \leq3000$, whereas, for \emph{LiteBIRD} $\ell_{\rm min} = 2$ and  $L_{\rm min} = 1$. We also present the constraint for $L\geq30$ in grey in order to provide a comparison with the other two experiments and indicate the effect that access to lower multipoles has on the QE constraint. The forecasted 95\% confidence limits are noted with dashed vertical lines. The 95\% upper limits are presented in table \ref{tab:Physical Constraints}.}
    \label{fig:likelihood}
\end{figure*}

We carry out the analysis described in section \ref{sec:likelihood} for BICEP/Keck, SO, and \emph{LiteBIRD}. The likelihood curves for $A_{CB}$ are depicted in Fig.~\ref{fig:likelihood}. We show the curves for $\beta = 1$ in order to draw easier comparisons with current constraints. Using these likelihood curves we calculated the 95\% upper limit on $A_{CB}$ and these are summarized in Table.~\ref{tab:Physical Constraints}. Below we discuss the salient features of the derived $A_{CB}$ constraints for each of the experiments we study in this work, comparing the results of the QE and BB techniques for each experiment.

\emph{BICEP/Keck}: The $A_{CB}$ limits presented by BICEP/Keck are derived using the QE method using a maximum CMB multipole of $l_{\rm max}=700$ \cite{Ade2017}. In our BICEP/Keck simulated forecast, on using the inputs from Table~\ref{tab:Instrument inputs} and evaluating the 95\% upper limits on $A_{CB}$ from the QE likelihood, we find that the resulting upper limit matches the limits found by BICEP/Keck, as seen in the top left panel of Fig.~\ref{fig:likelihood}. Specifically note that in our setup we have $l_{\rm max}=3000$ and the corresponding instrument beam and noise in the measurements naturally determine the weighting for the different modes.  Using an identical setting we also evaluate the BB likelihood and find that the  constraint on $A_{CB}$ is only 1.4 ($A_{\rm lens}=1$) times larger than the QE  constraint. Naturally we find that the BB constraint could be improved if the B-mode map were to be de-lensed (by a factor of 1.5 in the case of perfect de-lensing). However this finding contrasts the claim in \cite{Ade2017}, that constraints on $A_{CB}$ using the BB technique are \emph{significantly} worse than those derived using QE. In order to understand the origin of this contrasting conclusion, we also evaluate the BB likelihood using publicly available BICEP band power data\cite{BICEP2015_rlikelihood} and find a relatively weaker constraint on $A_{CB}=7.51 \times 10^{-5} {\rm rad}^{2}$, a factor of 2.6 worse than the QE constraint. This weakening of the constraints can be understood as a consequence of using a truncated multipole range in the BB likelihood analysis\footnote{BICEP used $l_{\rm max}=332$ for the BB likelihood, however it uses an $l_{\rm max}=700$ for evaluting their QE likelihood}. Finally we also checked that that using the $l_{\rm max}=332$ when carrying out the simulated forecast results in a constraint that is consistent with that derived from the BICEP band power data. These tests and checks allow us to benchmark our forecasting tools with published constraints hence allowing us to now perform reliable forecasts for upcoming experiments.
\par 
\emph{Simons Observatory}: As seen in both Figure~\ref{fig:likelihood} and Table~\ref{tab:Physical Constraints}, for the SO SAT the $A_{CB}$ constraints derived from the BB likelihood are comparable to those derived from QE. These constraints are similar to the current best upper limits from ACT of $A_{CB} =
1.0\times10^{-5}\,{\rm rad}^{2}$, derived using the QE method \cite{Namikawa2020}. However, for LAT, the QE analysis gives a significantly better constraint on $A_{CB}$ than BB and is also 2.6 times better than the QE limits achievable from the SAT.
\par 
De-lensing has a significant impact on the BB constraints achievable using the SAT, with perfect de-lensing yielding an improvement on BB constraints from the SAT by a factor of 1.76. For the LAT perfect de-lensing can also improve BB limits, however, the QE constraints continue to be better, as seen in Table~\ref{tab:Physical Constraints}. In summary, the best constraints for SO come from the QE analysis on the LAT. Therefore, in analogy to weak lensing, carrying out QE analysis on the LAT to place the best possible constraint on CB, is likely the best strategy for putting limits on CB induced B-mode power. It is possible to apply the QE analysis on de-lensed B-mode skies but we have not explored this analysis strategy\footnote{A careful treatment would require using revised noise estimates on the de-lensed B-mode maps.}. 
\par

\emph{LiteBIRD}:
\emph{LiteBIRD} is a space based experiment, and hence it significantly differs from both BICEP and SO/SAT, owing to its significantly larger sky coverage.

This immediately implies access to the largest angular scales on the sky, and this significantly enhances the constraints on $A_{CB}$ from  the QE likelihood, since the scale invariant CB power spectrum has more power at lower multipoles. Consequently we find that \emph{LiteBIRD} will be able to place constraints on $A_{CB}$ that are better than all other experiments considered in this work. 
\par The white noise is expected to be smaller for  \emph{LiteBIRD} than that of BICEP/Keck and SO, as seen in Table~\ref{tab:Instrument inputs}. 
Owing to this the constraints on $A_{CB}$ from the BB likelihood, even assuming no de-lensing, yields a constraint that is better than the best constraints achievable by any other experiment considered in this study. However, for \emph{LiteBIRD}, the constraints on $A_{CB}$ from the QE method yield constraints which are better by a factor of $\sim 16$ than those achievable from BB with perfect de-lensing. 

To better appreciate the gains from being able to use the large angle modes, we also derive $A_{CB}$ constraints that would be achievable by ignoring multipoles $L<30$ in the QE likelihood analysis and the resultant constraints are presented in Table~\ref{tab:Physical Constraints}. We find that the $A_{CB}$ constraints degrade by a factor of $\sim 30$ and become only marginally better than the QE constraints for SO LAT. We also carry out a similar exercise for the BB likelihood and find insignificant changes to the $A_{CB}$ constraints.

\begin{table}[h!]
    \centering
    \begin{tabular}{ccccc}
    \hline\hline
        & Instrument & $A_{CB}$ $[{\rm rad}^{2}]$ & $B_{1{\rm Mpc}}$ [nG] &  $g_{\phi\gamma\gamma}\,[H_{I}^{-1}]$\\
        \hline\hline
       QE & BICEP/Keck & $3.32\times10^{-5}$ & 30 & $7.24\times10^{-2}$\\
         
         & SO LAT & $3.96\times10^{-6}$ & 10.4 & $2.83\times10^{-2}$ \\
         
         & SO SAT & $1.03\times10^{-5}$ & 16.8 & $4.03\times10^{-2}$ \\
         
         & \emph{LiteBIRD} & $6.84\times10^{-8}$ & 1.4 & $4.65\times10^{-3}$ \\
        ($L > 30$) & \emph{LiteBIRD}  & $2.31\times10^{-6}$ & 8.0 & $1.90\times10^{-2}$ \\
         & ACT & $1.0\times10^{-5}$ & 16.6 (est.)& $4.0\times10^{-2}$\\
          \hline
        BB & BICEP/Keck & $4.65\times10^{-5}$ & 36 & $8.57\times10^{-2}$\\
        
        $A_{\rm lens} = 1$& BK data & $7.51\times10^{-5}$ & 46 & $1.09\times10^{-1}$\\
         
         & SO LAT & $1.63\times10^{-5}$ & 19.1 & $4.56\times10^{-2}$ \\
         
         & SO SAT & $1.25\times10^{-5}$ & 18.5 & $4.43\times10^{-2}$ \\
         
         & \emph{LiteBIRD} & $2.91\times10^{-6}$ & 9.0 & $2.14\times10^{-2}$ \\
         \hline
         BB-DL & BICEP/Keck & $4.24\times10^{-5}$ & 34 & $8.18\times10^{-2}$\\
         $A_{\rm lens} = 0.7$& SO LAT & $1.30\times10^{-5}$ & 18.9 & $4.53\times10^{-2}$ \\
         
         & SO SAT & $1.11\times10^{-5}$ & 17.5 & $4.18\times10^{-2}$ \\
         
         & \emph{LiteBIRD} & $2.46\times10^{-6}$ & 8.2 & $1.97\times10^{-2}$ \\
         \hline
         
         BB-NL & BICEP/Keck & $3.12\times10^{-5}$ & 29 & $7.01\times10^{-2}$\\
         $A_{\rm lens}=0$ & SO LAT & $1.26\times10^{-5}$ & 18.6 & $4.45\times10^{-2}$ \\
         & SO SAT & $7.12 \times 10^{-6}$ & 14.0 & $3.35\times10^{-2}$\\
         & \emph{LiteBIRD} & $1.12\times10^{-6}$ & 5.6 & $1.33\times10^{-2}$\\
         \hline\hline
    \end{tabular}
    \caption{Forecasted 95\% upper bounds on $A_{CB}$, which is converted into a limit on PMF field strength, with a reference frequency of $\nu = 150$ GHz, or a PNGB-photon coupling, for both QE and BB calculated with $\beta=1$. For BB this is carried out for $A_{\rm lens}=1$, $A_{\rm lens}=0.7$ (BB-DL) and $A_{\rm lens}=0.0$ (BB-NL). These constraints are calculated for BICEP/Keck, SO, and \emph{LiteBIRD}. The constraint from the analysis of the BICEP/Keck $B$-mode data is also presented (BK data). The values for ACT presented here are from \cite{Namikawa2020} except the constraint on $B_{1\rm Mpc}$, which is estimated from the ACT constraint on $A_{CB}$.}
    \label{tab:Physical Constraints}
\end{table}

\subsection{Constraints on Physical Phenomena}
While the focus till now has been on constraining CB, the origin and motivation for this work, the forecasted  constraints can be used to place constraints on a more broad range of phenomena which result in CPR.
\par 
One possible cause of CPR could be Faraday rotation induced by PMFs \cite{PhysRevD.86.123009,De2013,10.1093/mnras/stt2378}. The PMF amplitude is related to the CB power spectrum via the following relation \cite{PhysRevD.92.123509,Ade2017}:
\begin{equation}
    B_{1{\rm Mpc}} = 2.1\times10^{4}\,{\rm nG}\left(\frac{\nu}{30{\rm GHz}}\right)^{2}\sqrt{\frac{L(L+1)C^{\alpha\alpha}_{L}}{2\pi}}\;.
\end{equation}
Here, $\nu$ is the observed photon frequency. Assuming a scale invariant form for the CB power spectrum results in the following relation between the PMF strength and $A_{CB}$,
\begin{equation}\label{eq:ACB2B}
    B_{1{\rm Mpc}} = 2.1\times10^{4}\,{\rm nG}\left(\frac{\nu}{30{\rm GHz}}\right)^{2}\sqrt{A_{CB}}\;.
\end{equation}
Note that the above equation is independent of multipole and can be used to translate $A_{CB}$ constraints to those on the amplitude of the PMF. We invert Eq.~\ref{eq:ACB2B} at $\nu=150$ GHz to derive the upper limits on the primordial magnetic field strength for the different experiments and these are tabulated in Table \ref{tab:Physical Constraints}. Not surprisingly the constraint we forecast on the PMF strength, $B_{1 Mpc}$, using the QE method matched the limits presented by BICEP/Keck\cite{Ade2017}. 
\par
We find from the forecasts for the QE method that LiteBIRD could improve upon the existing BICEP/Keck limits by a factor of $\sim 21$. While there is no quoted constraint on PMF amplitude from ACT, we translate $A_{CB}$ constraints from ACT in \cite{Namikawa2020} using the same prescription as above and the resultant PMF constraint is given in  Table~\ref{tab:Physical Constraints}. Comparing this estimate to the QE forecast for \emph{LiteBIRD} suggests that \emph{LiteBIRD} may also be able to improve on any potential constraints ACT might place on $B_{1\rm Mpc}$ by nearly a factor of $\sim 12$. We find that the \emph{LiteBIRD} experiment is expected to give a PMF constraint that is a factor of $\sim 8$ better than the PMF constraint expected from SO LAT, and a factor of $\sim 12$ better than the PMF constraint expected from SO SAT.
\par

The primary motivation for this work, cosmological birefringence, is usually attributed to the coupling of a PNGB  to the gauge field. There are a variety of exciting and physically well motivated physical candidates for the PNGB including the dark matter axion, therefore constraining the strength of the coupling between the PNGB and the photon is of great interest. One can convert to an axion photon coupling, $g_{\phi\gamma\gamma}=\beta/2M$ and the relationship between $g_{\phi\gamma\gamma}$ and the CB power spectrum is \cite{Ade2017, Caldwell2011},
\begin{equation}
    g_{\phi\gamma\gamma} = \frac{4\pi}{H_{I}}\sqrt{\frac{ L(L+1)C^{\alpha\alpha}_{L}}{2\pi}}\;,
\end{equation}
or, according to \eqref{eqn:Birefringence Power spectrum} assuming the PNGB doesn't obtain a mass during inflation,
\begin{equation}
    g_{\phi\gamma\gamma} = \frac{4\pi}{H_{I}}\sqrt{A_{CB}}\;.
\end{equation} 
Here, $H_{I}$ is the Hubble parameter during inflation.
We have also used the forecasted constraints on $A_{CB}$ to place possible upper bounds on $g_{\phi\gamma\gamma}$ for Simons Observatory, and \emph{LiteBIRD} and these are presented in Table~\ref{tab:Physical Constraints}. As a test case, we compute an upper bound on $g_{\phi\gamma\gamma}$ using the QE method for BICEP/Keck and find that it matches the existing QE bound found in \cite{Ade2017}.
\par 
The best constraint on $g_{\phi\gamma\gamma}$ is expected to come, once again, from \emph{LiteBIRD}. Morever, the \emph{LiteBIRD} experiment could significantly improve upon existing bounds by nearly a factor of $\sim 16$ over the BICEP/Keck bound \cite{Ade2017} and nearly a factor of $\sim 9$ over the ACT bound \cite{Namikawa2020}. \emph{LiteBIRD} may give a constraint on $g_{\phi\gamma\gamma}$ that is a factor of $\sim 6$ better than the forecasted constraint from SO LAT and a factor of $\sim 9$ better than the forecasted constraint from SO SAT .

\subsection{Effect of birefringence on primordial B-modes}

\begin{figure*}[ht!]
    \centering
    \includegraphics[width=1\textwidth]{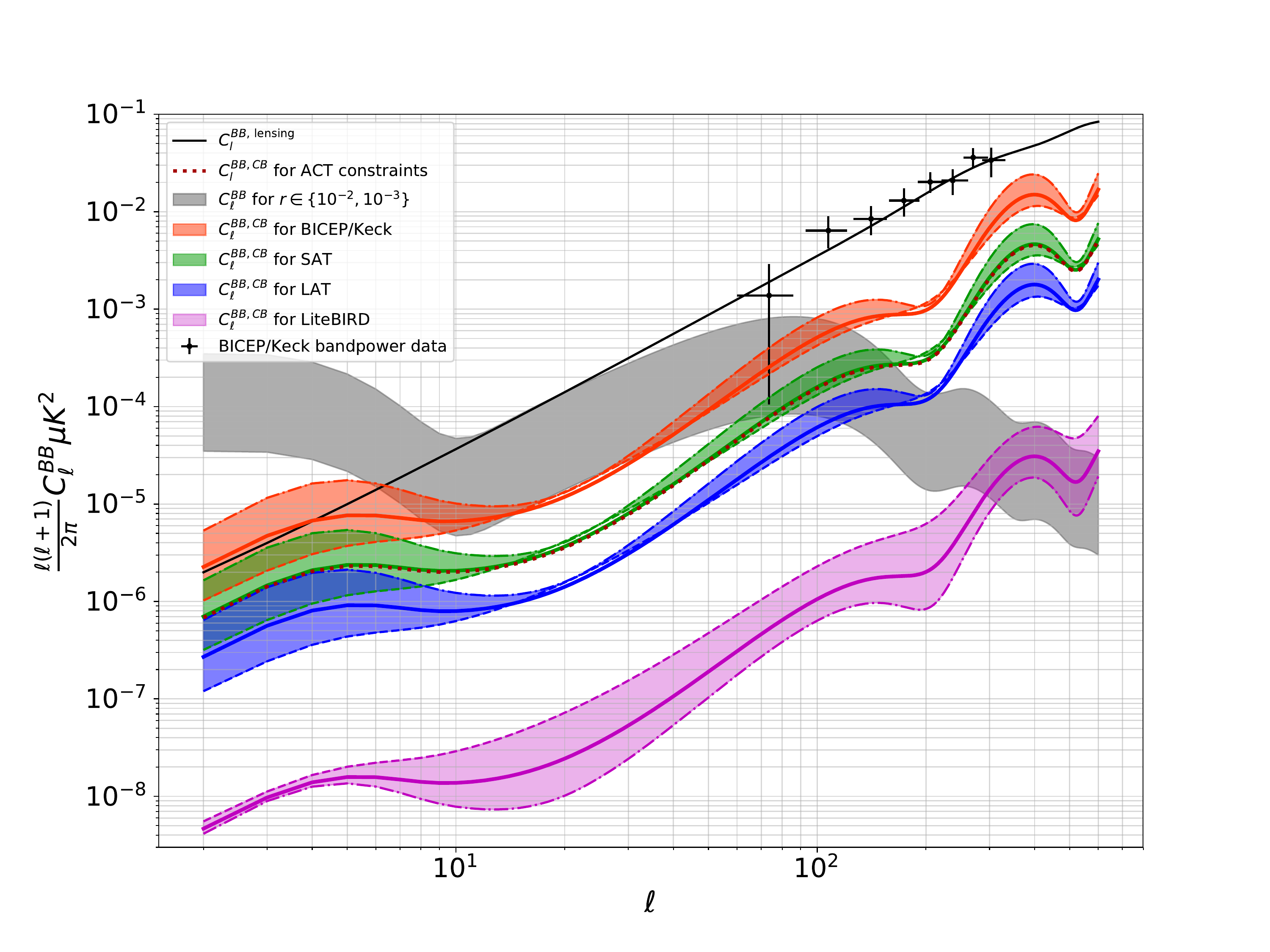}
    \caption{Maximum CB induced B-modes spectra, for varying CB spectral slopes: $C^{\alpha\alpha}_{L}\propto (L(L+1))^{-0.85}$ (dashed lines), scale-invariant $C^{\alpha\alpha}_{L}\propto (L(L+1))^{-1}$ case (solid lines) and $C^{\alpha\alpha}_{L}\propto (L(L+1))^{-1.15}$ (dot-dashed lines), with best $A_{CB}$ upper limits derived using the QE likelihood. The $C^{BB,CB}_{\ell}$ are compared to the primordial $\tilde{C}^{BB}_{\ell}$ when the tensor to scalar ratio $r$ is in the range $\sim 10^{-2} - 10^{-3}$. The CB induced $C_{l}^{BB,CB}$ allowed by the most recent ACT constraint is shown as a dotted maroon line.}
    \label{fig:BK primordial r comparison}
\end{figure*}

As seen in Eq.~\eqref{eqn:deltaclbb} additional $B$-mode power is induced by CB. We argued and demonstrated how this can be used to place constraints on CB from the measurements of $C^{BB}_{l}$. We now turn our attention to understanding the potential implications of the CB constraint study on measurement of primordial B-modes, since this is a primary science goal for the experiments considered in this work. Under the assumption that CB exists,  we characterize the contamination by evaluating the induced $B$-mode power spectrum from the best 95\% upper limits placed on $A_{CB}$ for all experiments. That is, we take the best 95\% confidence limits on $A_{CB}$ for each experiment, use them as the amplitude for CB spectra and compute the $B$-mode induced by each of these spectra. To gain some insight on the dependence on the slope of the CB spectra, we extend our analysis by estimating the QE likelihood constraints on $A_{CB}$, for CB spectra which deviate from the scale invariant form, specifically assuming $\beta=[0.85,1.15]$. Note that when deriving constraints on $A_{\rm CB}$ for these non-scale invariant spectra, we re-normalize $C_{L}^{\alpha}$ to match the scale invariant spectra amplitude at the pivot multipole $L_0=1$. While we do not provide these constraints here, we use these to evaluate the corresponding induced B-mode spectra, which are depicted in Fig.~\ref{fig:BK primordial r comparison}.
\par
Since the shape of the B-mode spectra induced by CB differs from that induced by tensors, we compare the amplitudes at $\ell\sim 80$.  For all the ground based experiments, the CB induced B-mode power spectra can constitute of order 1 to a few ten percent of the total primordial B-mode for $r\in[10^{-2},10^{-3}]$ as can be seen in Fig.~\ref{fig:BK primordial r comparison}. Further we note that the shallower CB spectrum ($\beta=-0.85$) results in a induced B-mode spectrum with a smaller amplitude while a steeper CB spectrum ($\beta=-1.15$) results in a B-mode spectrum with a larger amplitude. This, as we now argue, is quite counter intuitive. With pivoting at $L_0=1$ and $A_{\rm CB}=1$, it is clear that the $C_{\ell}^{BB}$ induced by a shallower CB spectrum will have an overall higher amplitude than one induced by a steeper CB spectrum, owing to greater amplitude of the shallower spectrum than the steeper spectrum at most multipoles (except at $L_0$). Since QE analyses are highly sensitive to low multipoles, on reducing access to the large angle modes, the $A_{\rm CB}$ constraints on a steeper CB spectrum are more weakened than they are for a shallower CB spectrum. This weakening of constraints results in the induced $C_{\ell}^{BB}$ from a steeper CB spectrum having a higher amplitude than that induced from the shallower CB spectrum, when evaluated for the best $A_{\rm CB}$ limits derived from the QE likelihood, resulting in the corresponding ordering of the induced B-mode spectrum - explaining the counter intuitive trend.

For \emph{LiteBIRD} the story is significantly different, owing to enhanced sensitivity and access to low multipoles, the $A_{CB}$ constraints are more stringent. Therefore the CB induced B-mode spectra have a very low amplitude and cannot act as major contaminants to measurements of primordial B-modes with the corresponding target amplitudes. Here it is interesting to note that the steeper CB spectrum induces a smaller B-mode spectrum than the shallower CB spectrum, as one may have expected. In this case, since the $A_{\rm CB}$ constraints are driven dominantly low multipoles, there is relatively smaller variation in the $A_{\rm CB}$ limits and the B-mode spectra induced by CB spectra of different slopes follow the expected trend.

\section{\label{Discussion}Discussion}

Future prospects for the detection and constraint of CB by the upcoming \emph{LiteBIRD} and SO experiments, and current constraints on CB from the contemporary BICEP/Keck experiment have been discussed. The constraints these experiments should be able to place on CB, or more generally CPR, using both a QE and BB approach have been compared. It was found that the QE approach yields the strongest constraints. However, it was also found that that in the SO SAT and BICEP/Keck forecast that the QE approach will yield only a marginal improvement over the constraint found using the BB method. In such cases, the computational expense and relative complexity of the QE method may make the use of the BB approach a much more favourable option for placing constraints on anisotropic CB. As long as the CB signal is found to be consistent with zero using BB, no QE investigation is needed. However, if in these cases a non-zero CB signal is found using the BB method, a QE method must be used to verify this signal is in fact sourced by CB.
\par The much weaker constraint from the BICEP/Keck band-power $B$-mode data compared to the strength of the constraint in the forecast BICEP/Keck case, along with the degradation in quality of the forecast constraint when the multipole range is limited, suggests that limits in multipole range may have lead to the lower quality BICEP/Keck constraint from BB analysis mentioned in \cite{Ade2017}. The strongest constraint is expected to come from the \emph{LiteBIRD} experiment using a QE approach. The \emph{LiteBIRD} is expected to offer roughly an order of magnitude improvement over the other experiments included in this analysis when using a QE approach. This improvement will be due to the improved sky coverage a low multipole access of the instrument. 
\par 
The constraints on CPR were also used to predict constraints on the possible physical sources of CPR, both primordial magnetic fields and cosmological birefringence. More specifically, we calculated upper bounds on both the field strength for primordial magnetic fields, and for the PNGB-photon coupling. Stronger constraints on $A_{CB}$ will yield stronger constraints on both the PNGB-photon coupling and PMF field strength, and it was found that \emph{LiteBIRD} is expected to be able to place the strongest constraint on both of these physical parameters.
\par
    In previous searches for primordial CMB $B$-mode the cosmological birefringence effect has been seen as largely irrelevant. The perspective has been that, while it is may be an interesting prospect to detect such an exotic physical effect, CB has little or no bearing on more mainstream CMB cosmology. However, as CMB experiments continue to probe primordial $B$-modes with more and more sensitivity, and as de-lensing techniques become more efficient CB may become a significant potential contaminant to $B$-mode science that needs to be excluded. In this work we have shown that the presence of an anisotropic CB effect with a power spectrum of an amplitude allowed by current upper bounds will indeed induce a large enough B-mode signal to act as a contaminant in future surveys seeking to constrain the tensor to scalar ratio in the range $r\sim10^{-2}$ to $r\sim10^{-3}$. Therefore, it will be necessary to truly ensure that such a contaminant is indeed not present. For future experiments such as SO it will be important to perform a QE based analysis to ensure that any detected $B$-mode signal is not from a CB contaminant. If a significant CB is detected, its influence on primordial B-mode studies can be removed by first de-rotating the polarization sky to remove the effects of CB, using techniques that bear strong resemblance to the de-lensing analysis \cite{Gluscevic2009,Kamionkowski2009}. Other extensions of these techniques are currently being studied with the aim of mitigating systematics induced B-mode power\cite{Williams2020}.  A detected CB signal would be an exciting prospect and a signal of new exciting physics. However, such a detection would warrant the removal of additional induced $B$-mode power in order to recover the true primordial $B$-mode signal.

\acknowledgments

JW was supported by Science and Technology Facilities Council (STFC) studentship. AR was supported by the ERC Consolidator Grant {\it CMBSPEC} (No.~725456) as part of the European Union's Horizon 2020 research and innovation program.


\bibliographystyle{JHEP}
\bibliography{bibliography.bib}
%\begin{thebibliography}{99}
%\bibliography{bibliography.bib}
%\end{thebibliography}
\end{document}